\documentclass[runningheads]{llncs}

\usepackage{amssymb,amsmath}
\usepackage{capt-of}
\usepackage{mathtools}
\usepackage{xspace}
\usepackage{listings}
\usepackage{anyfontsize} 
\usepackage{float}
\usepackage{graphicx}
\usepackage{subcaption}
\usepackage{tikz}
\usepackage{pgfplots}
\usepackage{colortbl}
\usepackage{relsize}

\usepackage[normalem]{ulem}  

\newcommand{\tata}{\textsc{Contata}\xspace}
\newcommand{\vamp}{\textsc{Vampire}\xspace}
\newcommand{\toolname}{\textsc{Cataclyst}\xspace}

\newcommand{\vars}[1]{\ensuremath{\textit{vars}(#1)}\xspace}

\newcommand{\Prs}{\ensuremath\Phi\xspace}
\newcommand{\Fn}{\ensuremath f\xspace}
\newcommand{\Msk}{\ensuremath{\mathcal M}\xspace}

\newcommand{\Tm}{\ensuremath{\mu}\xspace}
\newcommand{\Prim}{\ensuremath{\Lambda}\xspace}

\newcommand{\Sk}{\ensuremath{S}\xspace}

\newcommand{\N}{\ensuremath{\mathbb N}\xspace}

\newcommand{\nil}{\texttt{nil}\xspace}

\newcommand{\nilp}{\texttt{endp}\xspace}

\newcommand{\head}{\texttt{head}\xspace}
\newcommand{\tail}{\texttt{tail}\xspace}
\newcommand{\cons}{\texttt{cons}\xspace}
\newcommand{\ttt}[1]{\texttt{#1}\xspace}

\newcommand{\insrt}{\texttt{insert}\xspace}
\newcommand{\gov}[1]{\textit{gov}(#1)\xspace}

\makeatletter
\newcommand{\sx}[1]{%
  \text{%
    \ttfamily(%
    \@sxprocess#1 \@nil
    )%
  }%
}
\def\@sxprocess#1 #2\@nil{%
  #1%
  \if\relax\detokenize{#2}\relax
  \else
    \ %
    \@sxprocess#2\@nil
  \fi
}
\makeatother

\newcommand{\bool}{\ttt{bool}}
\newcommand{\hl}[1]{\ensuremath{_{(#1)}}}
\newcommand{\thl}[1]{\hl{#1}}

\newcommand{\None}{\ensuremath{\textit{None}}\xspace}

\newcommand{\Synth}{\ensuremath{\textit{synth}}\xspace}

\newcommand{\CspObj}{\ensuremath{\textit{CandidateSpace}}\xspace}
\newcommand{\Csp}{\ensuremath{\mathcal{U}}\xspace}
\newcommand{\CspNxt}{\ensuremath{\textit{\Csp.next(\ )}}\xspace}
\newcommand{\CspPr}[1]{\ensuremath{\textit{\Csp.prune(#1)}}\xspace}

\newcommand{\ConSp}{\ensuremath{\textit{con\_space}}\xspace}

\newcommand{\EmSp}{\ensuremath{\textit{em\_space}}\xspace}
\newcommand{\EmSps}{\ensuremath{\textit{em\_spaces}}\xspace}

\newcommand{\Cdt}{\ensuremath{\textit{cand}}\xspace}
\newcommand{\Cex}{\ensuremath{\textit{cex}}\xspace}
\newcommand{\Constraints}{\ensuremath{\textit{constraints}}\xspace}
\newcommand{\Cpt}{\ensuremath{\textit{concept}}\xspace}

\newcommand{\Genrr}{\ensuremath{\textit{generalize}}\xspace}
\newcommand{\Ver}[4]{\ensuremath{\Verr(#1,#2,#3,#4)}\xspace}
\newcommand{\Verr}{\ensuremath{\textit{check}}\xspace}

\DeclarePairedDelimiter{\angles}{\langle}{\rangle}
\DeclarePairedDelimiter{\dangles}{\langle\!\langle}{\rangle\!\rangle}

\DeclarePairedDelimiter{\abs}{\lvert}{\rvert}
\newcommand{\notlt}{\ensuremath{\not<}\xspace}

\newcommand{\FA}[2]{\ensuremath{\dangles{\forall #1 :: #2}}\xspace}
\newcommand{\EX}[2]{\ensuremath{\dangles{\exists #1 :: #2}}\xspace}

\newcommand{\Proph}{\textit{Proph}\xspace}
\newcommand{\Retro}{\textit{Retro}\xspace}
\newcommand{\Naive}{\textit{No-gen}\xspace}

\newcommand{\prefixminn}{\#59\xspace}

\begin{document}

\title{Recursive Program Synthesis from Sketches \\and Mixed-Quantifier Properties
}
\titlerunning{Synthesis from Sketches and MQ Properties}

\author{Derek Egolf \and Stavros Tripakis}
\authorrunning{D. Egolf and S. Tripakis}

\institute{Northeastern University, Boston, MA, USA\\
\email{egolf.d@northeastern.edu, stavros@northeastern.edu}}

\maketitle

\begin{abstract}
We present a novel approach for the synthesis of recursive programs from relational, mixed-quantifier first-order logic properties. We solve this problem using a sketching-based, enumerative, counterexample-guided approach. Our algorithm learns syntactic constraints from counterexamples to prune the candidate space and employs a prophylactic pruning technique to avoid enumerating invalid candidates altogether. We implement our technique in a tool called Cataclyst and evaluate it on a suite of 60 benchmarks. We demonstrate that both counterexample generalization and prophylactic pruning significantly improve performance. Cataclyst solves 59/60 benchmarks, while variants of Cataclyst without counterexample generalization or prophylactic pruning solve fewer benchmarks. No existing tool can directly handle the entire class of synthesis problems that our tool can. A portfolio of existing state-of-the-art tools solves 30/60 of our benchmarks.
\end{abstract}


\section{Introduction}

The synthesis of provably correct software from formal specifications has been a long-held aspiration of computer science~\cite{Church57,MannaWaldinger1980,PnueliRosner89,Finkbeiner2016NATO,GulwaniPolozovSingh2017}.
Although the field has come a long way since its early days, the automated synthesis of general programs from general logical specifications remains limited. In this paper, we focus on the synthesis of general {\em recursive functional programs} from general {\em mixed-quantifier} (MQ) first-order logic properties.

MQ properties arise naturally when specifying program correctness.
For example, suppose we want to synthesize a
function \ttt{(prefixb xs ys)} that decides whether list \ttt{xs} is a prefix of list \ttt{ys}.
The soundness of \ttt{prefixb} is naturally expressed by the MQ property
\[
    \FA{\ttt{xs}, \ttt{ys}}{\sx{\ttt{prefixb} \ttt{xs} \ttt{ys}}
    \implies\EX{\ttt{suffix}}{\ttt{ys} = \ttt{xs} \cdot \ttt{suffix}}}
\]

We tackle the problem using the sketching paradigm~\cite{sketch}. In
this paradigm, the user provides a {\em sketch}, which is a partial program
with holes, and the synthesis tool fills in the holes to produce a complete
program that satisfies the given specification. Sketching allows the user to
syntactically bias the space of candidate programs using their expert
knowledge. Sketching does not trivialize the synthesis problem. Typically, a
sketch still characterizes an infinite candidate space. Even if we bound the
size $n$ of the expressions used to fill the sketch, the candidate space
typically grows exponentially in $n$. 
And even verifying a single candidate solution against the specification is
hard (typically undecidable).

To address scalability issues, we adopt a novel grammar-guided, property-based testing-guided, and
counterexample-guided approach. We carefully enumerate the space of candidate
programs using a lazy, on-the-fly approach that avoids sending candidates to
the verifier as much as possible. When (and only when) we encounter a candidate
program that requires verification, we use state-of-the-art
property-based testers~\cite{cgen,z3} to search for a counterexample. Our algorithm learns
syntactic constraints on the candidate programs from each counterexample. These
syntactic constraints are used to prune the candidate space. Our pruning
technique is {\em prophylactic}: we make a best effort to avoid enumerating
pruned candidates altogether, rather than iterating over them and then
rejecting them. 
Our evaluation shows that our prototype tool \toolname can solve 59 out of 60 benchmarks, with each benchmark requiring no more than 2 mins. All 59 synthesized programs are proven correct, 39 of them fully automatically.

\section{Problem Definition}
\label{sec:prelims}

\subsection{Recursive Programs and Properties}

We use this running example throughout the paper. The task is to synthesize a
function $\insrt : \ttt{int} \times \ttt{int-list} \to \ttt{int-list}$, which
satisfies the {\em property} 
$\phi := \FA{\ttt x,\ttt y,\ttt{xs}} {\ttt y\in \ttt{xs}
\implies \ttt y \in \sx{\insrt \ttt x \ttt{xs}}}$.
I.e., inserting an element into a list should not remove any existing elements.
For this example, we also require that $\insrt$ satisfies the {\em test} (a
quantifier-free sentence) 
$\tau := \sx{\insrt \ttt 3 \sx{\ttt 1 \ttt 2}} = \sx{\ttt 1 \ttt 2
\ttt 3}$.
I.e., inserting \ttt 3 into the list \sx{\ttt 1 \ttt 2} should yield the list
\sx{\ttt 1 \ttt 2 \ttt 3}. Note that this unit test is specific to this example.
In general, the user is not required to provide unit tests, but such tests
might be useful, together with properties, in guiding the synthesis procedure
to the right solution.

An example of a synthesized implementation of $\insrt$ is given by the
following recursive definition, expressed in the LISP-like ACL2s programming
language~\cite{acl2s}:
\begin{lstlisting}[mathescape=true,basicstyle=\small\ttfamily,columns=fixed,basewidth=0.5em,escapechar=|]
(definec insert (x :int xs :int-list) :int-list
    (if (endp xs)
        (cons x nil)
        (cons (head xs) (insert x (tail xs)))))
\end{lstlisting}
This implementation places the element \ttt x at the end of the list \ttt{xs}
and leaves the other elements in their original order. Therefore, $\sx{\insrt
\ttt 3 \sx{\ttt 1 \ttt 2}}$ evaluates to \sx{\ttt 1 \ttt 2 \ttt 3}, satisfying
$\tau$. Furthermore, no elements are ever removed from the list,
so the synthesized \ttt{insert} also satisfies $\phi$.

We note that $\phi$ is defined using the predicate symbol $\in$. The
synthesized implementation also uses \head, \tail, \cons, and \nilp, as well as
the constant symbol \nil. Moreover, the type symbols \ttt{int} and
\ttt{int-list} are used. All of these symbols are part of a {\em background
theory} and in general we assume that the semantics of these symbols are fixed
before synthesis begins. We will write $\Prim$ to denote the set of function
symbols in the background theory. In contrast, we say that the function symbol
$\insrt$ is {\em under synthesis}.

Suppose there is one function symbol $f$ under synthesis and that the semantics
of all other symbols are given by a background theory $B$. Then if $e$ is an
implementation of $f$, we write $e \models_B \phi$ to mean that $e$ satisfies
the property $\phi$ under the semantics given by $B$. Formally, we can define
the meaning of $e \models_B \phi$ in terms of first-order logic
validity~\cite{harrison,fol-barrett,fol-ranise}. In particular, we use the
recursive function body $e$ to define the appropriate axioms for the symbol $f$
and extend $B$ with these axioms to obtain a new theory $B_e$. Then we say that
$e \models_B \phi$ if and only if $\phi$ is valid in $B_e$. First-order logic
validity and the details of this axiomatization are standard~\cite{acl2-car},
so we omit the details here. In general, we use $\vec f = f_1, \ldots, f_n$ to
denote a list of function symbols under synthesis, and $\vec e = e_1, \ldots,
e_n$ to denote their implementations. We write $\vec e \models_B \phi$ to mean
that together, the implementations $\vec e$ satisfy the property $\phi$ under
the semantics given by background theory $B$. In order for $\vec e \models_B
\phi$ to be coherent, we require that the implementations given by $\vec e$ are
{\em admissible} (c.f. Section~\ref{sec:admiss}).

Our logic and expression language are typed. We assume that the set of type
symbols and their meaning is fixed by the background theory $B$ before
synthesis begins (and that this set may be extended until synthesis begins). We
assume that all functions, both in the background theory and those under
synthesis, have a {\em function type} of the form $t_1 \times \cdots \times t_n
\to t$. Furthermore, we assume access to a countably infinite set of typed
variable symbols. Finally, we assume that each function symbol $f$ is
associated with a list of typed variable symbols $\vars{f}$ that is consistent
with its function type. For instance, the function symbol $\insrt$ has function
type $\ttt{int} \times \ttt{int-list} \to \ttt{int-list}$ and is associated
with the typed variable symbols $\ttt x : \ttt{int}$ and $\ttt{xs} :
\ttt{int-list}$. 

\subsection{Lazy Semantics and Governors}

When we evaluate a function like \insrt on a concrete input, only some
expressions are evaluated. For instance, when we evaluate $\ttt{(insert 3
\nil)}$,
we first evaluate the condition \ttt{(endp nil)}, which is true, and then we
evaluate the then-branch, but we do not evaluate the else-branch. I.e.,
\ttt{if} is evaluated {\em lazily}. We assume that all other function
applications are evaluated {\em eagerly}. I.e., we evaluate {\em all} arguments
to a function application, before evaluating the function application itself.

Let $e$ be an expression and let $e_\ell$ be a sub-expression of $e$ at
location $\ell$. Then the sub-expression $e_m$ at location $m$ in $e$ is a {\em
governor} of $e_\ell$ if the evaluation of $e_m$ governs whether $e_\ell$ is
evaluated. For instance, in the \insrt example, the expression \ttt{(endp xs)}
is a governor of the expression \ttt{(head xs)} because \ttt{(head xs)} is only
evaluated if \ttt{(endp xs)} evaluates to false. In this case, we say that
\ttt{(endp xs)} is a {\em negative} governor because it must evaluate to false
in order for \ttt{(head xs)} to be evaluated. A governor is {\em positive} if
it must evaluate to true to evaluate the governed expression. The predicate
$\gov \ell$ is the conjunction of all positive governors of $e_\ell$ and the
negation of all negative governors of $e_\ell$. $e_\ell$ is evaluated if and
only if $\gov \ell$ is satisfied. The set of positive and negative governors of
a sub-expression is computed by walking the if-statements and recording the
polarity of the conditions leading to the sub-expression.

\subsection{Admissibility}
\label{sec:admiss}

\paragraph{Contract Admissibility}

For background theory $B$, we assume that every background function $g\in\Prim$
is associated with an {\em input contract}. In particular, if $g$ has function
type $t_1\times\ldots\times t_n\to t$ then there is a predicate $p_g$ with type
$t_1\times\ldots\times t_n\to\bool$. I.e., the input contract $p_g$ is a
predicate on the inputs of $g$. Input contracts are used to characterize which
inputs are appropriate to pass to a background function. For instance, the
function $\ttt{head}$ returns the first element of a non-empty list; it is an
error to call $\ttt{head}$ on the empty list. Of course the background theory
may allow to interpret $p_g$ as a tautology, which is equivalent to omitting
the input contract.

A function under synthesis \Fn is contract admissible with respect to a
background function $g$ if whenever \Fn calls $g$, the inputs to $g$ are
guaranteed to satisfy the input contract of $g$. 
More precisely, $\Fn$ is contract admissible if for all locations $\ell$
in the body of \Fn where $g$ is called, $\gov \ell$ implies that the input 
contract of $g$ is satisfied.

For instance, in Fig.~\ref{fig:admiss-ex} function \ttt{no-admit} is
inadmissible with respect to \texttt{head} because the input contract of
\texttt{head} requires its input to be a non-empty list. The function
\ttt{yes-admit} in Fig.~\ref{fig:admiss-ex}, on the other hand, is admissible
with respect to \texttt{head} because the guard of the if-statement ensures
that \texttt{head} is only called with a non-empty list. 

\paragraph{Measure Functions}

Consider a function $\Fn : A\to B$. A {\em measure function} of $\Fn$ is a function $\Tm : A\to\N$ which is (strictly) {\em decreasing} with respect to $\Fn$. 
Suppose a recursive call $(\Fn\ \vec x')$ occurs at location $\ell$ in the body
of $\Fn$. Then $\Tm$ is decreasing with respect to $\Fn$ at location $\ell$ if
for all inputs $\vec x$ to \Fn, $\gov \ell$ implies that $\mu(\vec x') <
\mu(\vec x)$. $\Tm$ is decreasing with respect to $\Fn$ if it is decreasing at all
locations of recursive calls to \Fn. I.e., whenever the recursive call is
executed, the measure of the inputs to the recursive call decreases.

For instance, $\mu = \texttt{length}$ is not decreasing with respect to the
function \ttt{no-decr} in Fig.~\ref{fig:admiss-ex} because \texttt{no-decr} is
called with input \texttt{xs} in a recursive call, but $\texttt{(length xs)}
\notlt \texttt{(length xs)}$. In contrast, $\mu$ is decreasing with respect to
the function \ttt{yes-decr} in Fig.~\ref{fig:admiss-ex} because now the
recursive call input is \texttt{(tail xs)}, and $\texttt{(length (tail xs))} <
\texttt{(length xs)}$ for any non-empty list \texttt{xs}, and \texttt{xs} is
guaranteed to be non-empty in the recursive call to \texttt{yes-decr}. If a
measure is decreasing with respect to a function, we know that the function
terminates on all inputs. We note that the theory of measure functions
presented here can be extended to allow for a broader class of termination
proofs, but we omit the details~\cite{acl2-car,acl2s-termination}.

\begin{figure}
\centering
\begin{minipage}[t]{0.48\linewidth}
\begin{lstlisting}[mathescape=true,basicstyle=\fontsize{7.4}{9}\selectfont\ttfamily,columns=fullflexible,keepspaces=true,escapechar=|]
(definec no-admit (xs :int-list) :int
    (head xs))
(definec yes-admit (xs :int-list) :int
    (if (endp xs) 0 (head xs)))
\end{lstlisting}
\end{minipage}\hfill
\begin{minipage}[t]{0.48\linewidth}
\begin{lstlisting}[mathescape=true,basicstyle=\fontsize{7.4}{9}\selectfont\ttfamily,columns=fullflexible,keepspaces=true,escapechar=|]
(definec no-decr (xs :int-list) :int
    (no-decr xs))
(definec yes-decr (xs :int-list) :int
    (if (endp xs) 0 (yes-decr (tail xs))))
\end{lstlisting}
\end{minipage}
\caption{Examples of admissible (or not) functions.}
\label{fig:admiss-ex}
\end{figure}

Suppose $B$ is a background theory with background functions $\Prim$ and $\mu$
is a measure function for $f$. A recursive function definition of $f$ is {\em
admissible} with respect to $B$ and $\mu$ if it is contract admissible with
respect to all $g\in\Prim$ and if $\mu$ is decreasing with respect to $f$.

\subsection{Sketches and Grammars}

In addition to the signature of the function under synthesis and the properties
that this function must satisfy, our tool also takes as input a {\em sketch}
that places syntactic restrictions on the implementation of the function under
synthesis.
For instance, we may require that $\insrt$ is implemented by
filling the {\em holes} in the following {\em sketch body}:
\begin{lstlisting}[mathescape=true,basicstyle=\ttfamily,columns=fixed,basewidth=0.5em,escapechar=|]
(definec insert (x :int xs :int-list) :int-list
    (if |$h_1$| |$h_2$| (cons |$h_3$| (insert |$h_4$| |$h_5$|))))
\end{lstlisting}
As an example, we might obtain the following {\em completion} of the sketch body
\begin{lstlisting}[mathescape=true,basicstyle=\ttfamily,columns=fixed,basewidth=0.5em,escapechar=|]
(definec insert (x :int xs :int-list) :int-list
    (if (endp xs)|\hl 1|
        (cons x nil)|\hl 2|
        (cons (head xs)|\hl 3| 
              (insert x|\thl 4| (tail xs)|\hl 5|))))
\end{lstlisting}
by filling the hole $h_1$ with the expression \ttt{(endp xs)}, etc. For
clarity, we write the expression $e$ as $e\thl i$ to indicate that $e$ fills
the hole $h_i$. 

Each hole of a sketch $h_i$ is associated with a {\em non-terminal} from a {\em
grammar} $G$ that constrains the set of expressions that can be used to fill
the hole\footnote{In general, multiple non-terminals can be attached to a hole.
Then, a hole may be filled with an expression derived from any of its
non-terminals.}.
For instance,
\begin{align*}
    \ttt I &\to \ttt x\mid\sx{\head \ttt L} \\
    \ttt L &\to 
        \nil
        \mid\ttt{xs}
        \mid\sx{\cons \ttt I \ttt L}
        \mid\sx{\tail \ttt L} \\
    \ttt B &\to \sx{\nilp \ttt L} 
\end{align*}

is a grammar that can be used to fill the holes of the sketch body for
$\insrt$. Then, for example, we associate holes with non-terminals as follows:
$h_1 : \ttt B, h_2 : \ttt L, h_3 : \ttt I, h_4 : \ttt I, h_5 : \ttt L$. Notice,
for instance, that the expression \ttt{(endp xs)} can be derived from the
non-terminal \ttt B as: $\ttt B \to \sx{\nilp \ttt L} \to \sx{\nilp \ttt{xs}}$.
Therefore, it is appropriate to fill the hole $h_1$, which is associated with
the non-terminal \ttt B, with the expression \ttt{(endp xs)}. 

Our grammars are a variant of normal-form regular tree
grammars~\cite{tree-automata}, but without a designated start non-terminal
(since holes have their own start non-terminals). Formally, a grammar $G$ is a
tuple $\angles{\Sigma, \Prim, \Gamma, R}$, where $\Sigma$ is a set of
terminals, $\Prim$ is a set of function symbols, $\Gamma$ is a set of
non-terminals, and $R$ is a set of {\em rules}. A rule is an expression of the
form $N \to E$, where $N\in\Gamma$ is a non-terminal symbol and
$E\in\Sigma\cup(\Prim\times\Gamma^*)$ is either a terminal symbol from
$\Sigma$ or a function symbol from $\Prim$ applied to a tuple of non-terminal
symbols from $\Gamma$. 

In our setting, $\Sigma$ is a finite subset of the variable symbols given by
the background theory $B$ and $\Prim$ is the set of background functions of
$B$. Constant symbols are treated as nullary function symbols. Recall that
$\Prim$ does not contain any function symbols under synthesis.  We use the
traditional inductive definition to describe when an expression $e$ is {\em
generated} by a non-terminal $N$. In particular, $e\in\Sigma$ is generated by
$N$ if there exists a rule $N \to e$ in $R$. Otherwise, $(f\ e_1\ \ldots\ e_k)$
is generated by the non-terminal $N\in\Gamma$ if there exists a rule $N\to (f\
N_1\ \ldots\ N_k)$ in $R$ such that each $e_i$ is generated by $N_i\in\Gamma$.
We say that an expression $e$ is a {\em concept} of a grammar $G$ if there
exists a non-terminal $N\in\Gamma$ such that $e$ is generated\footnote{As
    discussed, the terminals and function symbols are typed by the background
    theory. We also assume that non-terminals are typed and that a non-terminal
can only generate expressions of its type.} by $N$.

Every function under synthesis is associated with one or more sketch bodies
(c.f. multi-sketches below).
For simplicity, we assume that the same grammar $G$ is used to fill
the holes of all sketch bodies\footnote{Separate grammars can always be merged
into a single grammar with appropriate renaming of non-terminals.}. Let $e_0$
be a sketch body for a function $f$ under synthesis and suppose the holes of
$e_0$ are $h_1, \ldots, h_k$ and that the non-terminal of $h_i$ is $N_i$. An
{\em emergent} $(h_1, e_1), \ldots, (h_n, e_n)$ is an association list mapping
holes to concepts, where $e_i$ is generated by $N_i$. The implementation $e$ is
a {\em completion} of the sketch $e_0$ if there exists an emergent $(h_1, e_1),
\ldots, (h_n, e_n)$ such that $e$ is obtained by substituting each $h_i$ with
$e_i$ in $e_0$. We assume that the grammar and sketch for $f$ are such that the
only variables appearing in any completion $e$ of the sketch are those in
$\vars{f}$.

\paragraph{Multi-sketches}
A function $f$ under synthesis may be associated with multiple alternative
sketch bodies. A {\em multi-sketch} $\Msk$ for $f$ is the collection of all
sketch bodies associated with $f$. An implementation $e$ is a completion of a
multi-sketch $\Msk$ if there exists some sketch body in $\Msk$ such that $e$ is
a completion of that sketch body. Intuitively, a multi-sketch represents a union
of sketches---any one of the sketch bodies in $\Msk$ can be used to obtain an
implementation for $f$.

\subsection{The Synthesis Problem}

\begin{problem}[Synthesis]
    \label{prob:multi}
    Let $B$ be a background theory and let $f_1,\ldots,f_n$ be (typed) function
    symbols under synthesis. Let $\Msk_i$ be a multi-sketch for each $f_i$ and
    let $G$ be a shared grammar for the sketch holes.
    
    Then, given a set of properties \Prs
    and a list of measure functions $\mu_1,\ldots,\mu_n$,
    the synthesis problem is to find, if one exists, a list of
    implementations $\vec e = e_1,\ldots,e_n$, one for each $f_i$
    such that for all $i\in 1\ldots n$,
    $e_i$ is admissible with respect to $B$ and $\mu_i$,
    $e_i$ is a completion of $\Msk_i$,
    and $\vec e \models_B \Prs$.
\end{problem}

We emphasize that the measure functions are provided by the user. In our
benchmarks, the measure functions are typically simple and natural, e.g.,
the length of a list or number of nodes in a tree. An interesting idea is
to automatically synthesize measure functions, but we leave this for future
work.

We allow \Prs to be a set of {\em mixed-quantifier} (MQ) properties, which are
formulas that contain both universal and existential quantifiers. We assume
that these formulas are in prenex normal form. Additionally, we require that
whenever a function under synthesis is applied to a variable (or an expression
containing a variable), that variable is universally quantified and no
existentially quantified variable comes before it in the quantifier prefix of
the property.
For example, 
$\FA{\ttt{xs}, \ttt{ys}}
    {\EX{\ttt{suffix}}{
    (\ttt{prefixb}\ \ttt{xs}\ \ttt{ys}) 
    \implies \ttt{ys} = \ttt{xs} \cdot \ttt{suffix}}}$
is a typical MQ property: the function \ttt{prefixb} is applied to universally
quantified variables, and the existentially quantified variable is used as a
certificate. We do not constrain the quantifier prefix of the property
    (c.f. \ttt{prefixmin} in Fig.~\ref{fig:mq-properties}).
The restriction is only on which variables a function under synthesis can be
applied to. This restriction is algorithmically relevant in
Section~\ref{sec:genr}.

\section{Our Synthesis Algorithm}
\label{sec:algo}

Our synthesis algorithm works by (carefully) enumerating candidate programs and
checking them against the specification. If a candidate program is incorrect, a
{\em counterexample generator} produces a counterexample, which
demonstrates that the program is incorrect. This suggests a {\em brute force}
algorithm: enumerate, check, repeat until no counterexamples are found.
However, this brute force approach is not feasible, as the space of candidate
programs is typically far too large. To address this challenge, our algorithm
also {\em generalizes} each counterexample, producing a constraint on the
syntax of candidate programs. As the algorithm explores candidates,
new counterexamples are found, and the set of syntactic constraints
becomes larger, ruling out more and more of the search space.

We now present our algorithm in more detail. We will first show how to
synthesize a single function; in Section~\ref{sec:multi-fn}, we briefly discuss
how to extend it to synthesize multiple functions. The algorithm is represented
by \Synth in Fig.~\ref{fig:synth-algo}. Suppose we want to synthesize a single
function $f$, where the semantics of all background functions are given by a
background theory $B$.
The algorithm takes as input background theory $B$, a measure function $\mu$, a
set of properties $\Prs$, a grammar $G$, and a multi-sketch $\Msk$ for $f$.

The algorithm begins by initializing, as a function of $G$ and $\Msk$, a
candidate space object \Csp, which represents the space of candidate programs
(line~\ref{ln:csp-init}). Then, it repeatedly queries \Csp for a program \Cdt
at each iteration of the algorithm (line~\ref{ln:csp-next}). If \Csp fails to
return a program (indicated by \None, line~\ref{ln:csp-exhausted}), then the
space has been exhausted, and the algorithm returns \None, indicating that
there is no solution to the synthesis instance. Otherwise, the algorithm uses a
counterexample generator \Verr to check if \Cdt is admissible with respect to
$B$ and $\mu$ and if it satisfies the properties in $\Prs$
(line~\ref{ln:cgen}). If \Verr returns \None, then \Cdt is returned
(line~\ref{ln:found-solution}). Otherwise, \Verr returns a counterexample \Cex.
The algorithm then uses the subroutine \Genrr to generalize the \Cex into
\Constraints (line~\ref{ln:genr}). These constraints are then used to prune
\Csp (line~\ref{ln:csp-prune}), and the algorithm begins the next iteration.

In the case where no counterexample is found, we perform a final check with the
automated theorem prover ACL2s~\cite{acl2s}. This final check returns either
\ttt{QED} if \Cdt can be proven correct automatically, or \ttt{Unknown} if the
automated proof attempt fails. In the latter case, we still return the
unfalsified candidate to the user so that they may attempt manual proof and
inspection. See Section~\ref{sec:evaluation} for details on post-hoc
verification.

\begin{figure}[H]
\centering
\begin{minipage}[t]{0.48\linewidth}
\centering
\scalebox{0.85}{\input{figs/top-algo}}
\captionof{figure}{Synthesis loop}
\label{fig:synth-algo}
\end{minipage}
\hfill
\begin{minipage}[t]{0.48\linewidth}
\centering
\scalebox{0.85}{\input{figs/enum-algo}}
\captionof{figure}{Candidate enumeration}
\label{fig:enum-algo}
\end{minipage}
\end{figure}

Our counterexample generator (\Verr, line~\ref{ln:cgen}) treats $\forall^*$-properties and MQ properties
differently. For $\forall^*$-properties, we use the state-of-the-art
counterexample generator~\cite{cgen} of the ACL2s theorem prover~\cite{acl2s}
to obtain a counterexample to the property. For MQ properties, we use the SMT
solver Z3~\cite{z3}. Our approach is modular with respect to the choice of
counterexample generator.

The remainder of this section describes the key subroutines of the algorithm:
candidate enumeration ($\Csp.\textit{next}$), counterexample generalization,
and pruning ($\Csp.\textit{prune}$).

\subsection{Lazy, On-the-fly Enumeration without Pruning}
\label{sec:brute}

A brute force algorithm can enumerate candidate programs in two phases: first
enumerate all concepts from the grammar $G$ (up to some size bound), and then
combine these to obtain all possible emergents. Recall that emergents are
tuples of concepts and candidate programs (sketch completions) are obtained by
substituting emergents into sketches from the multi-sketch $\Msk$. In
Fig.~\ref{fig:synth-algo}, the enumeration of candidates is performed
on line~\ref{ln:csp-next} by a candidate space object \Csp, which is
instantiated on line~\ref{ln:csp-init} as a function of the grammar $G$ and the
multi-sketch $\Msk$. Internally, \Csp maintains two objects: a concept space
and a list of emergent spaces (one for each sketch in $\Msk$). We now describe
each of these objects in turn and then describe how they can be combined to
obtain lazy, on-the-fly enumeration of candidate programs.

\paragraph{Concept Enumeration} 

First, note that the concept space is independent of the multi-sketch $\Msk$.
It depends only on the grammar $G$.  At a high level, we start by enumerating
terminal concepts (i.e., constants and variables) from the grammar. These seed
a pool of concepts that have been enumerated so far. Then, we combine concepts
from this pool in a bottom-up manner to obtain new concepts, which are also
added to the pool. Enumerating expressions from a grammar in this way is
standard, so we refer the reader to Appendix~\ref{appendix:concept-enum} for
more details. We also note that concepts are enumerated in a lazy, on-the-fly
manner, rather than by generating an explicit list of concepts all at once.
Such lazy enumeration is also fairly standard.

\paragraph{Emergent Enumeration}

Suppose we have an explicit list $L$ of concepts from the grammar $G$ and that
there is just one sketch $\Sk$ in $\Msk$. Let $L_h$ be the list of concepts in
$L$ that are generated by the non-terminal of the hole $h$; i.e., those
concepts in $L$ that are appropriate to use to fill $h$. To enumerate all
emergents for $\Sk$, we need only enumerate all tuples in $L_{h_1}\times \ldots
\times L_{h_k}$, where $h_1,\ldots,h_k$ are the holes in $\Sk$. To enumerate
completions, we simply substitute each tuple into $\Sk$. In general, there are
multiple sketches in $\Msk$, and we can apply this procedure to each sketch
independently and take a union. As for concept enumeration, we can use an
on-the-fly product construction to lazily enumerate $L_{h_1}\times \ldots
\times L_{h_k}$ for each sketch.

\paragraph{On-the-fly Candidate Enumeration}

Fig.~\ref{fig:enum-algo} shows how to lazily enumerate emergents from a lazily
enumerated list of concepts (rather than an explicit list). At a high level,
the method repeatedly tries to obtain a new emergent by querying each emergent
space and querying the concept space for a new concept when all emergent spaces
are exhausted under the current list of concepts. Line~\ref{ln:emsp-loop} loops
over each emergent space, and queries each one for a new emergent
(line~\ref{ln:emsp-next}). If an emergent is found, it is returned
(line~\ref{ln:emsp-return}). If no emergent is found, the concept space is
queried for a new concept. If the concept space is exhausted
(line~\ref{ln:consp-exhausted}), then the entire candidate space is exhausted,
and \None is returned. Otherwise, the new concept is added to each emergent
space (lines~\ref{ln:emsp-add-loop} and~\ref{ln:emsp-add}). Once a new concept
is added, the loop repeats and the emergent spaces have a chance to produce an
emergent with the new concept. The emergent spaces maintain a chain of lazy
products and they extend this chain with new lazy products whenever a new
concept is added. These new lazy products capture all emergents that use the
new concept along with the previously cached concepts. 

Although on-the-fly enumeration allows us to obtain a new candidate program
without generating an explicit list of programs all at once, this enumeration
strategy is still not sufficient in practice. We will now discuss how to learn
syntactic constraints from counterexamples to prune the search space, then we
return to enumeration strategies that take advantage of these constraints.

\subsection{Counterexample Generalization}
\label{sec:genr}

In this section, we show how to generalize counterexamples to obtain {\em
syntactic constraints} on candidate programs. A syntactic constraint is either
an atom of the form $h \neq e$ where $h$ is a hole and $e$ is an expression, or
a disjunction of such atoms. Counterexamples can witness various kinds of
violations: contract violations (i.e., the program is contract inadmissible),
measure violations (i.e., the program's measure is non-decreasing), or property
violations. We start by giving examples of each kind of violation and the
corresponding generalizations, and then we describe the generalization
algorithms formally.

\subsubsection{Examples of Generalization}

\paragraph{Contract Violation Example}

The following program is an (inadmissible) completion of a sketch with 5 holes;
subscript $(i)$ indicates the expression that fills hole $h_i$:
\begin{lstlisting}[mathescape=true,basicstyle=\ttfamily,columns=fixed,basewidth=0.5em,escapechar=|]
(definec insert (x :int xs :int-list) :int-list
    (if (endp xs)|\hl 1|
        (tail xs)|\hl 2|
        (cons (head xs)|\hl 3| (insert x|\thl 4| xs|\thl 5|))))
\end{lstlisting}
E.g., $h_1$ is filled with $\sx{\nilp \ttt{xs}}$ and $h_5$ is filled with
\ttt{xs}. This program is contract inadmissible because it violates the input
contract of \ttt{tail} at location $h_2$. The counterexample \sx{\ttt x \ttt
0}, \sx{\ttt{xs} \nil} witnesses this violation; it requires to evaluate
\ttt{(tail nil)}, violating the contract of \ttt{tail}, which requires a
non-empty input. In this case, we generalize the counterexample to obtain the
constraint $h_1 \neq \sx{\nilp \ttt{xs}} \lor h_2 \neq \sx{\tail \ttt{xs}}$.
Intuitively, this constraint follows from the fact that the contract violation
is triggered by the call to \ttt{tail} at $h_2$, and $h_1$ is the only governor
of this location. It follows that all completions that fill $h_1$ with
$\sx{\nilp \ttt{xs}}$ and $h_2$ with $\sx{\tail \ttt{xs}}$ will exhibit the
same contract violation.

Contract violations are especially useful for pruning expressions that involve
deep nestings of destructors like \ttt{tail}. For instance, if a completion
involves the expression $\sx{\tail \sx{\tail \sx{\tail \ttt{xs}}}}$, then a
contract violation will be triggered unless the guards check that all of
$\ttt{xs}$, $\sx{\tail \ttt{xs}}$, and $\sx{\tail \sx{\tail \ttt{xs}}}$ are
non-empty. The generalization procedure will automatically produce constraints
that rule out completions where these guards are missing.

\paragraph{Measure Violation Example}
Let $\mu(\ttt x, \ttt{xs}) = \sx{\ttt{length} \ttt{xs}}$ be the measure
function for \insrt. Then, ignoring the contract violation, the above program's
measure function is not decreasing. In particular, we would need to prove the
following property to show that the measure is decreasing: $\FA{\ttt
x,\ttt{xs}}{\neg \sx{\ttt{endp} \ttt{xs}}_{(1)} \Rightarrow
\mu(\ttt{x}_{(4)},\ttt{xs}_{(5)}) < \mu(\ttt{x},\ttt{xs})}$. Here we have
annotated the property with the locations of the arguments to the recursive
call to \insrt, as well as the location of the governor of this recursive call.
A counterexample to this property could assign \ttt 0 to \ttt x and the list
$\sx{\ttt{1 2 3}}$ to \ttt{xs}. Here we would generalize the counterexample to
obtain the constraint $h_1 \neq \sx{\nilp \ttt{xs}} \lor h_4 \neq \ttt x \lor
h_5 \neq \ttt{xs}$. The intuition here is that the measure property that was
violated refers to expressions in the $h_1$, $h_4$, and $h_5$ positions, as
demonstrated by the annotations.

\paragraph{Property Violation Example}

Suppose we want to synthesize \insrt so that it satisfies the property that
the inserted element is a member of the resulting list, i.e.,
$\FA{\ttt x,\ttt{xs}}{\sx{\ttt{member} \ttt x \sx{\insrt \ttt x \ttt{xs}}}}$.
The following admissible program violates this property (annotated with
subscripts indicating hole expressions as before):
\begin{lstlisting}[mathescape=true,basicstyle=\ttfamily,columns=fixed,basewidth=0.5em,escapechar=|]
(definec insert (x :int xs :int-list) :int-list
    (if (endp xs)|\hl 1|
        nil|\hl 2|
        (cons (head xs)|\hl 3| 
              (insert x|\thl 4| (tail xs)|\thl 5|))))
\end{lstlisting}
In particular, the violation is witnessed by the counterexample $\sx{\ttt x
\ttt 0}, \sx{\ttt{xs} \nil}$. Evaluating \ttt{(insert 0 nil)} gives \ttt{nil},
and \sx{\ttt{member} 0 nil} evaluates to false. We generalize this
counterexample to obtain the constraint $h_1 \neq \sx{\nilp \ttt{xs}} \lor h_2
\neq \nil$. Intuitively, this constraint follows from the fact that when we
evaluated \sx{\insrt \ttt 0 \nil}, we only had to evaluate the expressions in
the $h_1$ and $h_2$ positions. All completions that fill $h_1$ with $\sx{\nilp
\ttt{xs}}$ and $h_2$ with $\nil$ will exhibit the same property violation.

\subsubsection{Generalization Algorithms}

Although generalization varies based on the kind of violation, our algorithms
all follow a common pattern. They identify a set of holes that are
relevant to the violation exhibited by a candidate program and then return a
syntactic constraint that requires at least one of these holes to be filled
differently in future candidates. I.e., all generalization algorithms will
return a constraint of the form $\bigvee_{h\in H} h \neq e_h$ where $H$ is a
set of {\em relevant} holes and $e_h$ is the expression filling hole $h$ in an
incorrect candidate program $e$. 

The relevant holes are those that {\em intersect} with a set of relevant
program locations, and this set varies by violation type. A program location is
a pointer to a sub-expression of a candidate program. A program location $\ell$
intersects with a hole $h$ if $\ell$ points to a sub-expression of the
expression filling $h$ or if the expression filling $h$ is a sub-expression of
the expression at $\ell$.

\paragraph{Contract Violation Generalization}

The algorithm for generalizing contract violations takes an incorrect candidate
program $e$, the location $\ell$ of a contract violation in $e$, and the sketch
$\Sk$ from which $e$ was obtained. In the case of a contract violation, the
set of relevant program locations $L_\ell$ contains exactly $\ell$ and the
locations of the governors of $\ell$. The set of relevant holes $H_\ell$ is the
set of holes that intersect with a location in $L_\ell$. Finally, we return the
following syntactic constraint: $\chi_\textit{contract}(\Sk,e,\ell) :=
\bigvee_{h\in H_\ell} h \neq e_h$.

\paragraph{Measure Violation Generalization}

The algorithm for generalizing measure violations takes an incorrect candidate
program $e$, the location $\ell$ of a non-decreasing recursive call in $e$, and
the sketch $\Sk$ from which $e$ was obtained.  We return
$\chi_\textit{measure}(\Sk,e,\ell) := \bigvee_{h\in H_\ell} h \neq e_h$ where
$H_\ell$ is computed for $\ell$ in the same way as for contract violations.

\paragraph{Property Violation Generalization}

The algorithm for property violation takes as input an incorrect candidate
program $e$, a property $\phi$ that is violated by $e$, a counterexample $c$
witnessing the violation, and the sketch $\Sk$ from which $e$ was obtained. 
A counterexample $c$ is an assignment of values to the first block of
universally quantified variables in $\phi$. 
Since $c$ witnesses the violation, we must have
$e \not\models_B \phi_c$, where $\phi_c$ is the formula obtained by
substituting $c$ into $\phi$. If $\phi$ is a $\forall^*$-property, then
$\phi_c$ is variable-free and evaluates to false. If $\phi$ is an MQ
property, then $\phi_c$ is merely invalid (not valid) in the theory $B_e$.
Concretely, the counterexample generator falsified $\phi_c$ when it
returned $c$.

So, $\phi$ has an initial block of universally quantified variables $\vec x$
and $c$ assigns values to these variables. Moreover, the property $\phi$
contains one or more calls to the function $f$ being synthesized. We assume
that $f$ is admissible: there are no contract violations and $f$ is
terminating. For each call to $f$ in $\phi$, we evaluate the call\footnote{As
mentioned in Section~\ref{sec:prelims}, we assume that $\phi$ only applies
$f$ to variables occurring before the first existential quantifier. Because
$c$ is required to assign values to these variables, our algorithm can evaluate
these calls.}
using the values assigned to $\vec x$ by $c$. We then trace the execution of
$e$ on these calls, recording the program locations $L$ that are reached during
any of these executions. We then identify the set of holes $H$ that intersect
with a location in $L$. Finally, we return the following syntactic constraint:
$\chi_\textit{property}(\Sk,e,\phi,c) := \bigvee_{h\in H} h \neq e_h$ where
$e_h$ is the expression filling hole $h$.

\subsubsection{Correctness of Generalization}

Since the generalization algorithms are used to prune the candidate space, it 
is important that they do not {\em overprune}, i.e., eliminate any candidates that are correct. The
following theorem states that our generalization algorithms do not overprune.

\begin{theorem}
    Let $e$ and $e'$ be candidate programs obtained by filling holes in a
    sketch $\Sk$. If $e$ exhibits a contract violation (resp. measure
    violation) at location $\ell$, and $e'$ violates
    $\chi_\textit{contract}(\Sk,e,\ell)$ (resp.
    $\chi_\textit{measure}(\Sk,e,\ell)$), then $e'$ also exhibits a contract
    violation (resp. measure violation). If admissible $e$ violates a property $\phi$ with
    counterexample $c$, and admissible $e'$ violates
    $\chi_\textit{property}(\Sk,e,\phi,c)$, then $e'$ also violates $\phi$.
\end{theorem}

{\em Proof sketch:}
If $e'$ violates $\chi_\textit{contract}(\Sk,e,\ell)$, then $e'$ is going to
have the same expressions as $e$ at location $\ell$ and at the governors of
$\ell$. It follows that $e'$ will exhibit the same contract violation as $e$.
Likewise for $\chi_\textit{measure}(\Sk,e,\ell)$. 

For property violations, the argument is a bit more involved. Since $c$
witnesses the violation, we know $e \not\models_B \phi_c$. I.e., $\phi_c$ is
invalid in the theory $B_e$. Because the function symbol $f$ is admissible with
the body $e$, we can evaluate all calls to $f$ in $\phi_c$ using $e$ to obtain
a formula $\phi_e$. $\phi_e$ is invalid in the background theory since $\phi_c$
is. But we can also obtain $\phi_{e'}$ in the same way using $e'$ as the body
of $f$. Since $e'$ violates $\chi_\textit{property}(\Sk,e,\phi,c)$, $e'$ has
the same expressions as $e$ at all locations reached during the evaluation of
calls to $f$ in $\phi_c$. Therefore, $\phi_e$ and $\phi_{e'}$ will replace each
call to $f$ in $\phi_c$ with the same expressions, so $\phi_e$ and $\phi_{e'}$
are syntactically equivalent. Therefore, $\phi_{e'}$ is also invalid in the
background theory, and $e'$ also violates $\phi$.
\hfill$\Box$  

\subsection{Retrospective and Prophylactic Pruning}
\label{sec:enum}

We can incorporate the syntactic constraints learned from counterexample
generalization into the enumeration procedure described in
Section~\ref{sec:brute}. In particular, we can modify the procedure so that
emergents violating any constraint are skipped. We call the modification {\em
retrospective pruning}. In particular, when an emergent space is queried for a
new emergent, it checks if the next emergent satisfies the accumulated
syntactic constraints. If it does, that emergent is returned. If it does not,
that emergent is skipped, and a new emergent from the lazy product is obtained.
This filtering is repeated until either a valid emergent is found or the
emergent space is exhausted. We call this approach retrospective because in
order to identify that an emergent is invalid, we first have to iterate over it
and only then can we check if it satisfies the constraints.

A better approach is {\em prophylactic pruning}, which avoids iterating over
many invalid emergents in the first place. This approach exploits the fact that
in many cases, the syntactic constraints only mention a small number of holes.
This observation allows us to identify constraint violation after only filling
a subset of the holes in a sketch. For instance, suppose there are 20 holes and
a constraint mentions only holes $h_1$ and $h_2$. After filling $h_1$ and
$h_2$, we can check the partial emergent against the constraint. If the partial
emergent violates the constraint, we can skip filling the remaining 18 holes,
backtrack, and try a new combination of concepts for $h_1$ and $h_2$.

Recall that each emergent space maintains a chain of lazy products of concepts.
When a new syntactic constraint is learned and passed to an emergent space, the
emergent space in turn passes the constraint to each lazy product in its chain.
This allows each lazy product access to the list of syntactic constraints,
which allows to check partially constructed emergents. Lazy products are
standard objects, but it is challenging to implement them in a way that
supports prophylactic pruning through dynamic syntactic constraints.

A {\em constrained} lazy product $\Pi$ maintains a list of syntactic
constraints $\mathcal X$, a list $L$ of concepts, and an internal constrained
lazy product $\Pi'$, or $\Pi' = \bot$ as the base case. Additionally, it
maintains an index $i$ into $L$ that points to the current concept. Initially
$i$ is set to 0; $\Pi\textit{.reset()}$ sets $i$ back to 0 and calls
$\Pi'\textit{.reset()}$ if $\Pi'\neq\bot$. We can then enumerate emergents
that satisfy the constraints in $\mathcal X$ by implementing a method
$\Pi\textit{.next(prefix)}$. For the top-level call, $\textit{prefix}$ is the
empty tuple.

First, if the index $i$ is out of bounds, then $\Pi\textit{.next(prefix)}$
returns \None, indicating that the lazy product is exhausted. Otherwise, obtain
the current concept $e$ from $L$ using the index $i$. Then extend the prefix
with $e$ to obtain $\textit{prefix} \cdot e$. There are then two cases. 
(1) If $\textit{prefix} \cdot e$ does not violate any constraint in $\mathcal
X$, and $\Pi'=\bot$, then $\textit{prefix} \cdot e$ is returned.
(2) If $\textit{prefix} \cdot e$ does not violate any constraint in $\mathcal
X$, and $\Pi'\neq\bot$, then we return the result of
$\Pi'\textit{.next(prefix} \cdot e\textit{)}$ as long as it is not \None.

There are two additional cases that require {\em backtracking}.
(3) If $\textit{prefix} \cdot e$ violates a constraint in $\mathcal X$,
backtrack.
(4) If $\Pi'\textit{.next(prefix} \cdot e\textit{)}$ returns \None, backtrack.
Here, backtrack means increment the index $i$ to point to the next concept in
$L$, call $\Pi'\textit{.reset()}$ if $\Pi'\neq\bot$, and try again by calling
$\Pi\textit{.next(prefix)}$.

\subsection{Multi-function Synthesis}
\label{sec:multi-fn}

Our algorithm thus far handles the synthesis of a single function. No major
changes are required to handle multi-function synthesis. For purposes of
enumeration, it makes no difference if the emergents fill holes in one function
body or holes that are spread across multiple function bodies. Therefore, the
enumeration procedure need not change to accommodate multiple functions.
Generalizing from contract violations and measure violations is also unaffected
by the shift to multi-function synthesis because these violations are always
localized to a single function body. Generalizing from property violations in
the single-function setting requires us to obtain a trace from executing the
function on a counterexample input. The multi-function setting is identical
except that we may cross function boundaries during this execution and a
generalization may simultaneously constrain the contents of holes residing in
distinct functions. 

\section{Evaluation}
\label{sec:evaluation}

We implement our approach in a tool called \toolname and evaluate it on a suite
of 60 benchmarks. 
{\em Properties:} 9/60 benchmarks include both
MQ and $\forall^*$ properties, while 51/60 benchmarks
have $\forall^*$ properties only.
{\em Benchmark Sources:} 47/60 benchmarks are
adapted from the literature: \vamp~\cite{vampire-rec-synth} or
\tata~\cite{contata}. 8/60 benchmarks are new benchmarks that include MQ
properties; the remaining 5/60 benchmarks are new
$\forall^*$ benchmarks. Our
artifact will include all benchmarks and where they come from. A more detailed
breakdown of the benchmarks is included in
Appendix~\ref{sec:benchmark-details}. 

\toolname is written in Python and will be made publicly available and
submitted for artifact evaluation. Our experiments were run on a machine with
14 GB of RAM and an AMD Ryzen 7 7840U processor (3.3 GHz base clock). 

We
compare three variants of our tool: one that does not use pruning or
generalization (\Naive), one that uses retrospective pruning (\Retro), and one
that uses prophylactic pruning (\Proph). Generally, our implementation is
modular, and tool variants differ only in the
enumeration/pruning/generalization strategies used. \Retro and \Naive both use
the same on-the-fly, lazy candidate enumeration algorithm (c.f.
Section~\ref{sec:brute}), but \Retro does generalization (c.f.
Section~\ref{sec:genr}) and filters out candidates that violate the obtained
syntactic constraints (c.f. Section~\ref{sec:enum}). \Proph and \Retro both use
the same generalization procedure, but \Proph uses prophylactic pruning instead
of retrospective (c.f. Section~\ref{sec:enum}). 

Our tool requires the user to provide an upper bound on the size of the
expressions used to fill holes in the sketch, and each benchmark provides this
bound as a parameter. Our tool takes a measure function as input for each
synthesis task. For our benchmarks, the measure functions just compute the
literal size of the input to the synthesized function, e.g., the length of a
list or the number of atoms in an S-expression.

\begin{figure}
\begin{tikzpicture}
  \begin{axis}[
    width=0.78\linewidth,
    height=0.23\linewidth,
    scale only axis,
    xmin=0,
    xmax=900,
    ymin=0,
    xtick distance=100,
    ytick distance=10,
    xlabel={Timeout (s)},
    ylabel={Number of Solved Instances},
    grid=major,
    tick label style={font=\scriptsize},
    label style={font=\footnotesize},
    legend style={at={(0.99,0.03)}, anchor=south east,
      legend columns=1, font=\scriptsize},
  ]
    \addplot+[const plot, line width=1pt, solid, color=red, mark=none] coordinates {
      (3.84, 1)
      (4.68, 2)
      (5.42, 3)
      (7.85, 4)
      (8.89, 5)
      (9.28, 6)
      (10.25, 7)
      (11.14, 8)
      (11.61, 9)
      (13.15, 10)
      (13.80, 11)
      (15.23, 12)
      (16.25, 13)
      (16.58, 14)
      (16.63, 15)
      (17.18, 16)
      (17.65, 17)
      (18.16, 18)
      (18.16, 19)
      (18.20, 20)
      (18.58, 21)
      (19.01, 22)
      (19.25, 23)
      (19.70, 24)
      (20.07, 25)
      (20.61, 26)
      (21.52, 27)
      (21.77, 28)
      (23.75, 29)
      (24.01, 30)
      (25.05, 31)
      (26.12, 32)
      (27.48, 33)
      (27.72, 34)
      (27.85, 35)
      (29.42, 36)
      (30.95, 37)
      (34.43, 38)
      (38.27, 39)
      (39.71, 40)
      (39.89, 41)
      (40.10, 42)
      (41.60, 43)
      (41.74, 44)
      (44.89, 45)
      (44.92, 46)
      (46.86, 47)
      (48.17, 48)
      (49.57, 49)
      (49.89, 50)
      (51.01, 51)
      (56.14, 52)
      (60.46, 53)
      (68.53, 54)
      (72.32, 55)
      (82.68, 56)
      (84.02, 57)
      (98.18, 58)
      (108.98, 59)
      (109.98, 59)
      (900.00, 59)
    };
\addlegendentry{Proph}

    \addplot+[const plot, line width=1.5pt, dash pattern=on 0pt off 2.5pt, line cap=round, color=blue, mark=none] coordinates {
      (3.89, 1)
      (4.84, 2)
      (5.55, 3)
      (7.94, 4)
      (8.89, 5)
      (9.44, 6)
      (10.46, 7)
      (10.78, 8)
      (12.98, 9)
      (13.50, 10)
      (14.09, 11)
      (15.61, 12)
      (16.70, 13)
      (17.32, 14)
      (17.42, 15)
      (17.82, 16)
      (18.03, 17)
      (18.24, 18)
      (18.28, 19)
      (18.92, 20)
      (18.93, 21)
      (19.22, 22)
      (19.70, 23)
      (20.07, 24)
      (21.29, 25)
      (21.79, 26)
      (21.90, 27)
      (23.15, 28)
      (24.30, 29)
      (26.24, 30)
      (26.44, 31)
      (26.80, 32)
      (27.52, 33)
      (27.52, 34)
      (28.81, 35)
      (29.33, 36)
      (29.43, 37)
      (34.31, 38)
      (35.32, 39)
      (38.69, 40)
      (40.21, 41)
      (42.16, 42)
      (42.25, 43)
      (44.41, 44)
      (45.75, 45)
      (47.98, 46)
      (50.22, 47)
      (52.54, 48)
      (64.05, 49)
      (70.46, 50)
      (261.57, 51)
      (276.69, 52)
      (281.49, 53)
      (419.51, 54)
      (655.00, 55)
      (898.28, 56)
      (899.28, 56)
      (900.00, 56)
    };
\addlegendentry{Retro}

    \addplot+[const plot, line width=1pt, dotted, mark=none] coordinates {
      (3.92, 1)
      (4.85, 2)
      (7.83, 3)
      (8.32, 4)
      (10.17, 5)
      (11.68, 6)
      (12.09, 7)
      (13.01, 8)
      (14.65, 9)
      (17.15, 10)
      (17.19, 11)
      (19.84, 12)
      (21.46, 13)
      (24.66, 14)
      (24.71, 15)
      (25.21, 16)
      (27.95, 17)
      (28.71, 18)
      (29.24, 19)
      (29.29, 20)
      (29.90, 21)
      (31.35, 22)
      (31.42, 23)
      (31.53, 24)
      (32.46, 25)
      (32.67, 26)
      (45.37, 27)
      (56.37, 28)
      (76.20, 29)
      (77.18, 30)
      (91.70, 31)
      (94.16, 32)
      (96.84, 33)
      (99.92, 34)
      (101.34, 35)
      (101.61, 36)
      (107.70, 37)
      (114.31, 38)
      (117.20, 39)
      (120.65, 40)
      (123.60, 41)
      (131.48, 42)
      (135.24, 43)
      (144.61, 44)
      (173.14, 45)
      (198.45, 46)
      (218.61, 47)
      (226.04, 48)
      (261.69, 49)
      (789.73, 50)
      (790.73, 50)
      (900.00, 50)
    };
\addlegendentry{No-gen}
  \end{axis}
\end{tikzpicture}
\caption{Cactus plot comparing our tool variants using no generalization (\Naive),
retrospective pruning (\Retro), and prophylactic pruning (\Proph) on 60
benchmarks.
}
\label{fig:cactus}
\end{figure}

\subsubsection{Benchmark Results}

For a timeout of 15 minutes (900 seconds) per benchmark, \Proph solves 59/60
benchmarks, \Retro solves 56/60, and \Naive solves 50/60. The set of benchmarks
solved by \Proph is a superset of those solved by \Retro, which is a superset
of those solved by \Naive. The cactus plot in Fig.~\ref{fig:cactus} shows the
number of benchmarks $y$ that a variant can solve if it is allowed to run each
benchmark for $x$ seconds. For instance, there are 58 benchmarks that \Proph
can solve within 100 seconds, but for \Retro and \Naive there are 50 and 34
benchmarks, respectively, that they can solve within 100 seconds. Per-benchmark
results are included in Appendix~\ref{sec:full-benchmark-table}.

\subsubsection{Benchmarks with Existential Quantifiers}

\begin{figure}
\centering
\normalsize
$\displaystyle
\begin{aligned}
& \FA{\ttt{xs},\ttt{ys}}{\EX{\ttt{suffix}}{\ttt{(prefixb xs ys)}
\implies \ttt{ys} = \ttt{xs}\cdot\ttt{suffix}}} \\
& \FA{\ttt{xs},\ttt{ys}}{\EX{\ttt{prefix}}{\ttt{(suffixb xs ys)}
\implies \ttt{ys} = \ttt{prefix}\cdot\ttt{xs}}} \\
& \FA{\ttt{x},\ttt{xs}}{\EX{\ttt{i}}{\ttt{(contains x xs)} \implies \ttt{x} =
\ttt{xs}[\ttt{i}]}} \\
& \FA{\ttt{xs}}{\EX{\ttt{ys}}{\ttt{(not (endp xs))} 
\implies \ttt{xs} = \ttt{ys} \cdot \ttt{(lst-last xs)}}} \\ 
& \FA{\ttt{xs}}{\EX{\ttt p,\ttt s}{\ttt{(monotonic xs)}
\implies \ttt{xs} = \ttt{0}^{\ttt{p}}\cdot\ttt{1}^{\ttt{s}}}} \\
& \FA{\ttt{x},\ttt{xs}}{\EX{\ttt{i}}{\FA{\ttt{j}}{
\ttt{(prefixmin x xs)}
\\
&\ \ \ \ \ 
\implies (\ttt{x} = \ttt{xs}[\ttt i] \land (\ttt j < \ttt i \implies \ttt{x} < \ttt{xs}[\ttt j]) )}}} \\
& \FA{\ttt{d},\ttt{xs}}{\EX{\ttt{i},\ttt{j}}{\ttt{(not (threshold-list d xs))}
\\
&\ \ \ \ \ 
\implies (\ttt{i} < \ttt{j} \land \ttt{j} < \abs{\ttt{xs}} 
\land \abs{\ttt{xs}[\ttt{i}] - \ttt{xs}[\ttt{j}]} < \ttt{d})}} \\
& \FA{\ttt{tr},\ttt{k}}{\EX{\ttt{p}}{(\ttt{k} \geq \ttt 0 \land \ttt{(has-k-path tr k)})
\\
&\ \ \ \ \ 
\implies (\ttt{(is-path p tr)} \land \ttt{(length p)} = \ttt{k})}} \\
& \FA{\ttt{tr},\ttt{n}}{\EX{\ttt{p}}{\ttt{(has-path-sum tr n)} 
\\
&\ \ \ \ \
\implies (\ttt{(is-path p tr)} \land \ttt{(path-sum p)} = \ttt{n})}}
\end{aligned}$
\caption{Examples of MQ properties from our benchmark suite.}
\label{fig:mq-properties}
\end{figure}

Fig.~\ref{fig:mq-properties} shows examples of MQ properties from our
benchmark suite. 
\ttt{prefixb} and \ttt{suffixb} check if one list is a prefix or suffix of another, respectively. 
\ttt{contains} checks if an integer is a member of a list.
\ttt{lst-last} returns the last element of a list.
\ttt{monotonic} checks if a bit-vector is non-decreasing. 
\ttt{prefixmin} checks if an integer is the minimum of some prefix of a list. 
\ttt{threshold-list} checks if the absolute difference of every pair of elements in a list is above a given threshold. 
\ttt{has-k-path} checks if there is a path of length $k$ in a given tree.
\ttt{has-path-sum} checks if there is a path with a given sum in a given tree.

\subsubsection{Post-synthesis
Verification}

In 39/59 benchmarks solved by \Proph, the ACL2s theorem prover was able to
automatically 
prove
that the synthesized program satisfies the specification.
In 
the remaining 20 cases,
we manually proved the correctness of the synthesized program in ACL2s. (Most
proofs were relatively easy, and required only the addition of helper lemmas
and/or the specification of an induction scheme. Some proofs used a non-trivial
proof strategy.)
In summary,
all 59 programs synthesized by \Proph were proven correct, either automatically or manually,
which demonstrates that attempting to falsify candidate solutions during synthesis (instead of trying to fully verify them) is effective in producing correct solutions.
(We remark that verifying that a candidate recursive program satisfies a first-order logic property is undecidable.)

\subsubsection{Comparison with \vamp and \tata}
\label{sec_vampire}

For the sake of completeness, we also experiment with \tata~\cite{contata} and
\vamp~\cite{vampire-rec-synth}, although neither of these
two tools is directly comparable to ours. Neither \tata nor \vamp
supports sketches, and \tata does not support MQ properties. \vamp
does not support properties that relate two function applications involving
different arguments (e.g. $\sx{f x y} = \sx{f y x}$ or $\sx{rev \sx{rev x}}
= \ttt x$), and it does not support the synthesis of multiple functions if
those functions do not take the same arguments. 

\vamp is an automated theorem prover for first-order logic with
equality~\cite{vampire-main} with support for automatic
induction~\cite{vampire-ind20,vampire-ind21}. \cite{vampire-rec-synth}
describes an approach for using \vamp to synthesize recursive programs from
first-order specifications. This approach is implemented as an extension of
\vamp, and it is publicly available~\cite{vampire-rec-synth-tool}.

Since \vamp does not support synthesis from the entire class of properties that
appear in our benchmarks, we are only able to encode 40/60 of our benchmarks in
its input language. 
Since \vamp also does not take sketches as input, this information
is also not encoded in the input to \vamp.
When running \vamp on the 40/60 of our benchmarks that it is able to handle, we observe 6 distinct behaviors: (1) solves benchmark, (2)
times out, (3) out of memory, (4) segfault, (5) search fails to find a solution or prove that
none exists, or (6) exhibits a bug. In case 6, \vamp reports that it has found
a solution, but the purported solution merely asserts the declarative,
non-executable correctness specification. An example is provided in
Appendix~\ref{appendix_vampire}. In private communication, the authors
of~\cite{vampire-rec-synth} confirmed that this is a bug in \vamp. 

To give \vamp the best possible chance of success, we run four variants of
\vamp on our benchmarks. Two of these variants use an old version of \vamp from
around the time of the publication of~\cite{vampire-rec-synth}, as suggested by
the authors of~\cite{vampire-rec-synth} in a private correspondence. The other
two variants use a more recent version of \vamp in a default portfolio mode and
in a mode that is specialized for induction. All segfaults that we observed
occurred when using the old version of \vamp, and the bug described above
occurred in both the old and new versions. We count a benchmark as solved if
any of \vamp's four variants solves it within 15 minutes. In total, across the
four variants, \vamp solves 11/60 benchmarks. One of these 11 benchmarks is an
MQ benchmark, and the other 10 benchmarks have only $\forall^*$ properties.

These results are not surprising, since \vamp is required to produce a proof
object {\em before} constructing a candidate program. In contrast, our tool
enumerates candidate programs independently of proof search, and only attempts
proofs for fully instantiated candidates after they pass counterexample
generation. 
These results indicate that inductive, counterexample-guided, sketch-based
synthesis is more effective than deductive, proof-guided, non-sketch-based
synthesis.

\tata~\cite{contata} is a tool which synthesizes recursive programs from $\forall^*$-properties. 
We run \tata on the 51/60 benchmarks in our suite that contain only $\forall^*$ properties.
\tata times out in 23 cases after 15 minutes, and runs out of memory in 2 cases.
Overall, \tata solves 26/51 
$\forall^*$
benchmarks and is not applicable to the 9 MQ benchmarks.
A portfolio composed of \tata and the four \vamp configurations solves 30/60
benchmarks. This assumes that all 5 solvers are run in parallel with a timeout
of 15 minutes.

\subsubsection{Detecting Unrealizability}

If the user provides an input with no solution, we say their input is {\em
unrealizable}. It is possible, for instance, that the user provides a sketch
that has no completion that satisfies the specification. As mentioned, our tool
takes as input a size bound on the expressions used to fill holes. Therefore,
\toolname can halt and say ``unrealizable'' if it exhausts the search space. If
the user receives such a response, they know to either relax the size bound,
modify the sketch, or investigate whether any properties are contradictory. 

We converted 60 realizable benchmarks into 60 unrealizable (to our knowledge)
benchmarks by modifying the inputs to exclude known solutions. To evaluate our
tool's ability to detect unrealizability, we ran \Proph on these 60
unrealizable benchmarks with a 15-minute timeout. \Proph recognizes
unrealizability in 54/60 cases. \Naive and \Retro recognize unrealizability in
27/60 and 38/60 cases, respectively. When a variant failed to recognize
unrealizability, it timed out after 15 minutes. Neither \tata nor \vamp is
applicable to these benchmarks because they do not take sketches and the
unrealizability is a consequence of the sketch in these benchmarks.

\section{Related Work}
\label{sec:related}

\paragraph{Recursive Program Synthesis from Examples}

All of~\cite{smyth,escher,para,myth,lambda-sq,ChoL25} synthesize recursive programs
from input-output examples. Our work differs from these works in that we
synthesize programs from general first-order specifications. Of these works,
only \textsc{Smyth}~\cite{smyth} supports user-provided sketches.
\textsc{Para}~\cite{para} internally uses templates and fills them with
recursion-free expressions, but these templates 
are not provided by the user. 

\paragraph{Recursive Program Synthesis from $\forall^*$-Properties}

All of~\cite{contata,burst,synquid,leon,NeoPLDI2018,ItzhakyPPRS21} synthesize recursive programs from
properties. To our knowledge, these works do not support mixed-quantifier
specifications, but only $\forall^*$ properties.
Furthermore,~\cite{burst,synquid,leon,NeoPLDI2018,ItzhakyPPRS21} place restrictions on the type of
universally quantified specifications that can be used; they cannot express,
for instance, $\FA{\ttt x,\ttt y}{\sx{\ttt{add} \ttt x \ttt y} = \sx{\ttt{add}
\ttt y \ttt x}}$, nor can they express $\FA{\ttt{xs}}{\sx{\ttt{rev}
\sx{\ttt{rev} \ttt{xs}}} = \ttt{xs}}$. Except for~\cite{leon},
none of these works support {\em user-provided} sketches.
~\cite{ItzhakyPPRS21} focuses on synthesizing auxiliary recursive functions in
the context of heap-manipulating programs and separation logic.
Moreover, while~\cite{leon} takes a sketch {\em body} as input,
it does not take user-provided grammars for the holes as input.
Although these tools support less expressive properties, this
restriction may ease the burden of automatically verifying
synthesized programs.

In addition to the major differences mentioned above, our approach differs
algorithmically from those of~\cite{contata,burst,synquid,leon}. \cite{synquid}
uses refinement types to specify and synthesize recursive programs through
top-down decomposition. \cite{leon} combines deductive synthesis rules with
enumerative search to synthesize recursive programs from pre/post-conditions.
\cite{burst} synthesizes recursive programs in a bottom-up fashion by
exploiting angelic semantics and iteratively strengthening the specification.
\cite{contata} synthesizes programs from general universally quantified
specifications using constraint-annotated tree automata.
~\cite{ItzhakyPPRS21} uses deductive synthesis techniques
whereas we use {\em counterexample-guided, inductive} synthesis techniques.
The backtracking approach of~\cite{NeoPLDI2018} is similar to
ours.~\cite{NeoPLDI2018} explores the space of candidate programs by
constructing derivations from a grammar. When they backtrack, they ``undo''
derivation steps. Our algorithm, on the other hand, backtracks by discarding
the expression used to fill a hole in the sketch. I.e., we explore a Cartesian
product of hole-filling expressions, while~\cite{NeoPLDI2018} explores a tree
of derivations.

\paragraph{Recursive Program Synthesis from Mixed-Quantifier Properties}

To our knowledge, the only work that can in principle synthesize recursive
programs from MQ properties is \vamp~\cite{vampire-rec-synth} (c.f.
Section~\ref{sec_vampire}). At a high level, \vamp uses a deductive synthesis
approach~\cite{MannaWaldinger1980} which consists in synthesizing a function
$f(\vec x)$ that satisfies $\phi$ by searching for a constructive proof of
$\FA{\vec x}{\EX{y}{\phi[\vec x, y]}}$. (\cite{vampire-non-rec-synth} uses a
similar approach but synthesizes non-recursive functions.) Although in
principle \vamp can handle MQ properties, it was only able to solve 1/9 of our
MQ benchmarks, and was only able to solve 10/51 of our $\forall^*$ benchmarks:
see Section~\ref{sec_vampire} for a detailed comparison and lessons learned.

\paragraph{Syntax-Guided Synthesis}
Syntax-Guided Synthesis (SyGuS)~\cite{sygus-main} is a framework for program
synthesis that uses grammars to restrict the search space of candidate
programs. The syntax of the SyGuS Language Standard (version
2.1)~\cite{sygus-lang} allows grammars that admit recursive function
definitions. Despite this, none of the 6000+ publicly available SyGuS
competition benchmarks~\cite{sygus-bench} involve recursive
functions---recursion is never explicitly enabled, nor is it implied by the
presence of a function name in its own grammar. Furthermore, the
state-of-the-art SyGuS solver CVC5~\cite{cvc4sy,cvc5} does not currently
support the synthesis of recursive functions~\cite{cvc5-recursion-github}
(although it does support synthesis of non-recursive functions when recursive
functions are in the background theory),
which is why we cannot run CVC5 on our benchmarks.
We remark that we view our tool as complementary to a tool like CVC5: for instance, we can imagine a portfolio solver which runs CVC5 on non-recursive benchmarks and our tool on recursive benchmarks.
(On a similar note, none of~\cite{bool-fn-synth2021,kuncak2010,DQFMT} support
the synthesis of recursive functions.)

Both our work and~\cite{cvc4sy} enumerate concepts from a grammar. A key
difference between our work and~\cite{cvc4sy} is that we then enumerate {\em
combinations of concepts} (emergents) in order to fill a sketch. An
optimization employed by~\cite{cvc4sy} is to discard equivalent concepts during
enumeration. For instance, if $x+y$ has been enumerated, then $y+x$ is
redundant and they only use $x+y$ (and not $y+x$) as a building block for
larger concepts. We would refer to this kind of optimization as {\em concept
reduction}, as it avoids iterating over redundant concepts. In contrast, our
prophylactic pruning technique focuses on {\em emergent reduction}, which
avoids iterating over {\em emergents}. 

\paragraph{LLMs}

In this paper we propose a symbolic, LLM-free solution to the recursive program
synthesis problem. Comparing our tool with LLMs or neurosymbolic tools is
beyond the scope of this paper. Note that a symbolic tool can run for free on a
laptop. In contrast, a proprietary LLM runs on prohibitively expensive,
powerful hardware in the cloud, and is accessed through paid API calls. Despite
this disparity, the preliminary study~\cite{fmcad2026} suggests that symbolic
tools can match or exceed LLM performance on some class of inputs.

\section{Conclusions}

We propose a novel method for synthesizing recursive programs from
mixed-quantifier (MQ) specifications. Our tool, \toolname, solves 59/60
benchmarks in under 2 mins each, while a portfolio of state-of-the-art,
existing tools solves just 30/60 of our benchmarks when allowed a timeout of 15
minutes. We acknowledge that this comparison is apples-to-oranges, since the
constituent tools of the portfolio do not support sketching. 

Our algorithm learns syntactic constraints from counterexamples to prune the
candidate space, and we use a prophylactic technique to avoid enumerating
pruned candidates altogether. Both our counterexample generalization and our
prophylactic pruning techniques significantly improve the performance of our
algorithm.


\section*{Acknowledgment}
This material is partly supported by the National Science Foundation under Graduate Research Fellowship Grant \#1938052. This work has also been partly supported by NSF's FMitF (Formal Methods in the Field) program, under awards 2319500 and 2525087. Any opinions, findings, and conclusions or recommendations expressed in this material are those of the author(s) and do not necessarily reflect the views of the National Science Foundation.

\bibliographystyle{splncs04}
\bibliography{bib,auto}

\clearpage
\appendix
\section{Benchmark Table}
\label{sec:full-benchmark-table}

\begingroup
\setlength{\tabcolsep}{1.4pt}
\setlength{\arraycolsep}{1.4pt}
\begin{table}[H]
\centering
\resizebox{\columnwidth}{!}{%
\begin{tabular}{r|l|r|r|r|r|r|r|r}
\hline
\textbf{\#} & \textbf{Benchmark} & \textbf{Holes} & \textbf{NTs} & \textbf{Rules} & \textbf{Sol. Size} & \textbf{\Proph} & \textbf{\Retro} & \textbf{\Naive} \\
\hline
1 & vampire-tree-max & 12 & 3 & 7 & 24 & 51.01 & 52.54 & 789.73 \\
2 & tree-k-path-smt & 8 & 5 & 10 & 15 & 41.60 & 42.25 & 261.69 \\
3 & prefixb-smt & 8 & 5 & 8 & 16 & 30.95 & 35.32 & 226.04 \\
4 & monotonic-smt & 10 & 5 & 7 & 18 & 24.01 & 26.80 & 218.61 \\
5 & vampire-prefix-given-suffix & 7 & 3 & 7 & 12 & 48.17 & 64.05 & 198.45 \\
6 & tree-sum-path-smt & 6 & 5 & 11 & 17 & 56.14 & 70.46 & 173.14 \\
7 & tree-forest-leaves & 8 & 7 & 18 & 13 & 44.92 & 44.41 & 144.61 \\
8 & bst-insertion & 13 & 6 & 13 & 24 & 49.89 & 50.22 & 135.24 \\
9 & ext-nested-list-eq & 8 & 4 & 7 & 17 & 23.75 & 24.30 & 131.48 \\
10 & ext-mirror-ternary2 & 6 & 3 & 9 & 11 & 44.89 & 47.98 & 123.60 \\
11 & ternary-tree-depth & 7 & 4 & 15 & 11 & 39.71 & 40.21 & 120.65 \\
12 & ext-mirror-ternary & 6 & 3 & 9 & 11 & 39.89 & 42.16 & 117.20 \\
13 & zip-same-len & 7 & 5 & 11 & 15 & 98.18 & 29.33 & 114.31 \\
14 & nested-list-eq & 8 & 4 & 7 & 17 & 16.58 & 17.32 & 107.70 \\
15 & mirror-ternary2 & 6 & 3 & 9 & 11 & 27.48 & 28.81 & 101.61 \\
16 & vampire-division & 10 & 2 & 6 & 14 & 40.10 & 38.69 & 101.34 \\
17 & mirror-ternary & 6 & 3 & 9 & 11 & 26.12 & 27.52 & 99.92 \\
18 & ext-list-eq & 8 & 4 & 7 & 17 & 25.05 & 26.24 & 96.84 \\
19 & tree-forest-size & 8 & 7 & 18 & 12 & 46.86 & 45.75 & 94.16 \\
20 & list-eq & 8 & 4 & 7 & 17 & 21.52 & 23.15 & 91.70 \\
21 & evens-odds & 8 & 3 & 6 & 12 & 34.43 & 34.31 & 77.18 \\
22 & prefixmin-smt & 8 & 5 & 9 & 16 & 20.61 & 21.90 & 76.20 \\
23 & vampire-list-max & 8 & 3 & 7 & 16 & 27.85 & 27.52 & 56.37 \\
24 & pairs & 6 & 6 & 10 & 16 & 11.14 & 13.50 & 45.37 \\
25 & even-odd & 6 & 2 & 7 & 10 & 20.07 & 19.70 & 32.67 \\
26 & ext-mirror-tree2 & 5 & 3 & 8 & 9 & 19.70 & 20.07 & 32.46 \\
27 & ext-mirror-tree & 5 & 3 & 8 & 9 & 19.01 & 19.22 & 31.53 \\
28 & even-odd2 & 6 & 2 & 7 & 10 & 19.25 & 18.92 & 31.42 \\
29 & mirror-tree2 & 5 & 3 & 8 & 9 & 18.16 & 18.28 & 31.35 \\
30 & threshold-smt & 13 & 8 & 12 & 23 & 29.42 & 29.43 & 29.90 \\
31 & list-sort & 7 & 4 & 9 & 17 & 17.18 & 18.03 & 29.29 \\
32 & binary-tree-depth & 6 & 4 & 12 & 9 & 21.77 & 21.79 & 29.24 \\
33 & mirror-tree & 5 & 3 & 8 & 9 & 15.23 & 15.61 & 28.71 \\
34 & vampire-assoc-add & 5 & 2 & 6 & 7 & 27.72 & 26.44 & 27.95 \\
35 & split-on & 8 & 6 & 9 & 16 & 17.65 & 17.82 & 25.21 \\
36 & index-of-contains-smt & 6 & 5 & 9 & 11 & 18.16 & 21.29 & 24.71 \\
37 & suffixb-smt & 6 & 5 & 8 & 10 & 18.20 & 18.24 & 24.66 \\
38 & list-removal2 & 8 & 4 & 7 & 15 & 18.58 & 18.93 & 21.46 \\
39 & list-removal & 8 & 4 & 7 & 14 & 16.63 & 17.42 & 19.84 \\
40 & vampire-len-concat & 7 & 3 & 6 & 11 & 13.80 & 14.09 & 17.19 \\
41 & vampire-square-root & 7 & 2 & 4 & 12 & 16.25 & 16.70 & 17.15 \\
42 & list-rev-twice & 4 & 4 & 8 & 9 & 10.25 & 10.46 & 14.65 \\
\hline
\end{tabular}%
}
\caption{Benchmark comparison results (continued on the next page)}
\label{tab:benchmark_comparison_full}
\end{table}

\clearpage
\begin{table}[H]
\centering
\addtocounter{table}{-1}
\resizebox{\columnwidth}{!}{%
\begin{tabular}{r|l|r|r|r|r|r|r|r}
\hline
\textbf{\#} & \textbf{Benchmark} & \textbf{Holes} & \textbf{NTs} & \textbf{Rules} & \textbf{Sol. Size} & \textbf{\Proph} & \textbf{\Retro} & \textbf{\Naive} \\
\hline
43 & parenthesis & 6 & 4 & 10 & 10 & 13.15 & 12.98 & 13.01 \\
44 & vampire-subtraction & 6 & 2 & 5 & 9 & 8.89 & 8.89 & 12.09 \\
45 & list-insertion3 & 5 & 3 & 8 & 10 & 11.61 & 10.78 & 11.68 \\
46 & list-insertion & 5 & 3 & 8 & 10 & 9.28 & 9.44 & 10.17 \\
47 & list-insertion2 & 5 & 3 & 8 & 10 & 7.85 & 7.94 & 8.32 \\
48 & vampire-last-element-smt & 5 & 3 & 6 & 10 & 5.42 & 5.55 & 7.83 \\
49 & vampire-double & 3 & 2 & 4 & 5 & 4.68 & 4.84 & 4.85 \\
50 & sized-list & 1 & 3 & 5 & 6 & 3.84 & 3.89 & 3.92 \\
\hline
51 & suffixb-full & 12 & 7 & 10 & 20 & 84.02 & 898.28 & TO \\
52 & index-of-contains-full & 12 & 6 & 11 & 22 & 49.57 & 655.00 & TO \\
53 & trie-eq & 10 & 5 & 11 & 21 & 108.98 & 419.51 & TO \\
54 & ext-list-cmp & 12 & 4 & 9 & 25 & 41.74 & 281.49 & TO \\
55 & list-cmp & 12 & 4 & 9 & 25 & 38.27 & 276.69 & TO \\
56 & binary-tree-eq & 10 & 5 & 11 & 21 & 72.32 & 261.57 & TO \\
57 & monotonic-full & 22 & 7 & 9 & 40 & 82.68 & TO & TO \\
58 & prefixb-full & 14 & 6 & 9 & 26 & 68.53 & TO & TO \\
59 & prefixmin-full & 20 & 6 & 11 & 38 & 60.46 & TO & TO \\
60 & ternary-tree-eq & 12 & 5 & 12 & --- & TO & TO & TO \\
\hline
\end{tabular}%
}
\caption{Benchmark comparison results (continued)}
\label{tab:benchmark_comparison_full_continued}
\end{table}

\endgroup

The table is sorted by descending \Naive times, and then by
descending \Retro time when \Naive times out. In cases where \Naive doesn't
time out, \Proph and \Retro have similar performance. But when \Naive times
out, \Proph is significantly faster than \Retro. For instance, in one case
(prefixmin-full, \prefixminn), \Proph solves the benchmark in 60 seconds, while
\Retro times out after 15 minutes. {\em Holes} shows the number of holes in the
sketch, {\em NTs} shows the number of non-terminals in the grammar, {\em Rules}
shows the number of production rules in the grammar, and {\em Sol. Size} shows
the size of the emergent used to complete the sketch, i.e., the sum of the
sizes of the expressions used to fill all the holes of the sketch. The size of
an expression is 1 if it is a variable or constant. Otherwise, if the
expression is of the form \sx{f $e_1 \ldots e_n$}, then its size is 1 +
size($e_1$) + \ldots + size($e_n$).

\section{Benchmark Details}
\label{sec:benchmark-details}

In total we have 60 benchmarks in our suite. 37 of these are based on those
of~\cite{contata}, 10 are based on those of~\cite{vampire-rec-synth}, and 13
are new benchmarks that we created. Below we provide details about the
benchmarks in our suite.

\subsubsection{\tata Benchmarks}
37 of our benchmarks are based on those of~\cite{contata}. We note though
that~\cite{contata} does not handle sketches, nor mixed-quantifier
specifications, so the benchmarks are not identical, nor are the tools directly
comparable. For each of the 30 benchmarks of~\cite{contata}, we include a
corresponding benchmark (with added sketch information) in our suite. In 7 of
these 30 cases, we extended a~\cite{contata} benchmark with at least one
additional property and kept both the original and extended benchmarks in our
suite. 

\subsubsection{\vamp Benchmarks}
10 of our benchmarks are based on those of~\cite{vampire-rec-synth}. We note
that~\cite{vampire-rec-synth} does not handle sketches, so the benchmarks are
not identical, nor are the tools directly comparable. Only one of these is a
mixed-quantifier benchmark.

\subsubsection{New Benchmarks}
Our suite also includes 8 new benchmarks that use mixed-quantifier properties.
Finally, we include 5 additional benchmarks that are Skolemized variants of
some mixed-quantifier benchmarks. I.e., these require to synthesize both a
target function and one or more witness-generating functions for the Skolem
symbols introduced when eliminating the existential quantifiers.

\section{Concept Enumeration in Detail}
\label{appendix:concept-enum}

Initially, the concept space enumerates {\em terminal concepts} (i.e.,
constants and variables) from the grammar using {\em terminal rules} of the
form $N\to a$, where $a$ is a terminal symbol or constant. A cache tracks each
concept that has already been enumerated, along with the non-terminal(s) used
to generate each. Once all terminal rules have been exhausted, the concept
space enumerates a new concept by first selecting a rule of the form $N\to (f\
N_1\ldots N_k)$ from the grammar. It then draws, if possible, a concept for
each $N_i$ from the cache, and applies the function $f$ to obtain a new
concept. If it is not possible to draw a combination of concepts for the $N_i$
in such a way that obtains a new concept, the concept space repeats the process
with a different rule. If {\em none} of the rules can be used to obtain a new
concept, the concept space indicates that it has been exhausted.

As a simple example, suppose the grammar contains the rules $\ttt E \to \ttt x$
and $\ttt E \to \sx{\ttt + \ttt E \ttt E}$. Our strategy first enumerates the
concept $\ttt x$, using the first rule (a terminal rule). Then, using the
second rule, it would enumerate the following concepts: 
$\sx{\ttt + \ttt x \ttt x}$,
$\sx{\ttt + \ttt x \sx{\ttt + \ttt x \ttt x}}$,
$\sx{\ttt + \sx{\ttt + \ttt x \ttt x} \ttt x}$, 
$\sx{\ttt + \sx{\ttt + \ttt x \ttt x} \sx{\ttt + \ttt x \ttt x}}$, etc. 
Notice in this case that when the cache contains \ttt x, there is only one way
to combine concepts. But when the cache also contains $\sx{\ttt + \ttt x \ttt
x}$, there are 4 ways to combine concepts. In general, the number of ways to
obtain a new concept is the product of the number of cached concepts for each
non-terminal in the selected rule. 

A naive implementation of this enumeration procedure might maintain, for each
rule, something like an explicit list of all appropriate combinations of cached
concepts and then extend it (careful to avoid duplicates) whenever a new
concept is added to the cache. This approach is inefficient. Instead, we use an
on-the-fly product construction to lazily enumerate these combinations for each
rule. We construct new lazy products every time a new concept is added to the
cache, and we lazily chain the products together so that we can obtain a stream
of concepts without maintaining an explicit list.

\section{Details on \vamp bug and an example}
\label{appendix_vampire}

\vamp makes a distinction between functions allowed in the body of the
synthesized program and functions used only to specify properties. The
latter are labeled ``uncomputable.'' 
For instance, to specify the correctness of $\ttt{prefixb}$ in \vamp, we would 
write the following:

\begin{verbatim}
(declare-fun spec (lst lst Bool) Bool)

(assert (forall ((x lst) (y lst) (b Bool)) 
  (= (spec x y b)
  (and
    (forall ((z lst))
      (=> (= y (append x z)) b))
    (=> b
        (exists ((z lst))
          (= y (append x z))))))))
   
(set-option :uncomputable (spec))
\end{verbatim}

The code above defines the correctness predicate \texttt{spec} such that $\texttt{(spec x y b)}$ is
true iff the Boolean flag \ttt{b} correctly indicates whether list \ttt{x} is a prefix of list \ttt{y}.
This is done by the \ttt{assert} statement which defines   \ttt{(spec x y b)} to be equal to 
\begin{eqnarray*}
&  (\forall \ttt{suffix} :: \ttt{y} = \ttt{x}\cdot\ttt{suffix} \implies \ttt{b}) \\
&  \land \\
&  (\ttt{b} \implies \exists \ttt{suffix} :: \ttt{y} = \ttt{x}\cdot\ttt{suffix}) 
\end{eqnarray*}
which is a conjunction of (nearly)\footnote{Note that the $\exists$ can be
pulled out front in the second conjunct.} prenex normal form formulas that
is equivalent to
\begin{eqnarray*}
&  \ttt{b} \iff \exists \ttt{suffix} :: \ttt{y} = \ttt{x}\cdot\ttt{suffix}
\end{eqnarray*}
That is, \ttt{(spec x y b)} says that \ttt{b} is true iff \ttt{x} is a prefix of \ttt{y}.

When we ask \vamp to synthesize a function that satisfies \ttt{spec}, \vamp outputs
\begin{verbatim}
   (prefixb x y) := (spec x y true)
\end{verbatim}
which violates the requirement that \ttt{spec} is uncomputable and thus not allowed to be used in the body of the synthesized program.
In essence, \vamp does not synthesize an executable program; it simply parrots
the correctness predicate.

We remark that our tool \toolname avoids this problem by excluding specification-only functions from the grammar. Because these functions are not part of the grammar, by construction they cannot be part of the synthesized program expressions.

\end{document}